\documentclass[aps,preprint,showpacs,preprintnumbers,amsmath,amssymb]{revtex4}
\usepackage{amsmath,mathrsfs,amsbsy,color,graphicx,bm,amsthm,amsfonts}
\usepackage{units}
\usepackage{bbm}
\usepackage{times}
\usepackage{dcolumn}
\usepackage{mathrsfs}
\usepackage{amsmath,amssymb,epsfig}
\usepackage{amsmath}
\newcommand{\udots}{\mathinner{\mskip1mu\raise1pt\vbox{\kern7pt\hbox{.}}
\mskip2mu\raise4pt\hbox{.}\mskip2mu\raise7pt\hbox{.}\mskip1mu}}
\begin{document}

\title{Nonseparability of multipartite systems in dilaton black hole}
\author{Shu-Min Wu$^1$\footnote{Email: smwu@lnnu.edu.cn}, Xiao-Wei Teng$^1$, Wen-Mei Li$^1$, Yu-Xuan Wang$^1$, Jianbo Lu$^1$\footnote{Email: lvjianbo819@163.com} }
\affiliation{$^1$ Department of Physics, Liaoning Normal University, Dalian 116029, China}


\begin{abstract}
We investigate the nonseparability of N-partite quantum systems by employing the Abe-Rajagopal (AR) $q$-conditional entropy for both free bosonic and fermionic fields in the background of a Garfinkle-Horowitz-Strominger (GHS) dilaton black hole.
An intriguing finding  is that the Hawking effect of the black hole can generate a net  nonseparability of W state for fermionic field. Notably, we observe that fermionic  nonseparability exhibits a stronger robustness than its bosonic counterpart, while fermionic coherence is found to be weaker than bosonic coherence within the dilaton black hole background. Additionally, our analysis reveals that the nonseparability of GHZ state is more pronounced than that of W state, yet quantum coherence of GHZ state is comparatively weaker than that of W state in dilaton spacetime. These results suggest that choosing the appropriate quantum resources for different particle types and quantum state configurations is essential for effectively tackling relativistic quantum information tasks.
\end{abstract}

\vspace*{0.5cm}
\maketitle
\section{Introduction}
From Einstein's perspective, the gravitational collapse of sufficiently massive stars leads to the formation of black holes, fascinating and profound entities that challenge and enrich our understanding of the universe. In recent years, both indirect and direct astronomical observations have provided strong evidence confirming the existence of these mysterious cosmic objects \cite{Q3,Q4,Q5,Q6,Q7,Q8,Q9,Q10,Q11}. A prominent example of a black hole within this framework is the Garfinkle-Horowitz-Strominger (GHS) dilaton black hole, a solution that arises from the intriguing predictions of string theory \cite{rfv01,rfv02,rfv1,rfv2,rfv3}. The presence of a dilaton field can significantly alter the properties of black hole geometries, making the study of dilaton black holes crucial for both gravitational theory and quantum mechanics.
Due to their immense distances from us and their extraordinary properties, black holes have long remained enveloped in mystery, becoming a central focus of cutting-edge research. One of the most intriguing challenges in this field is the information loss paradox of the black hole \cite{Q12,rfv4,rfv5,rfv6}. As we all know, Hawking radiation is believed to be the root cause of the information loss paradox, as it suggests that information falling into a black hole may be irretrievably lost, seemingly violating the principles of quantum mechanics.  Quantum properties, particularly in the context of Hawking radiation, may offer a crucial pathway toward resolving this paradox. Moreover, in realistic environments, quantum systems are inevitably influenced by gravitational effects, making it essential to consider the impact of Hawking radiation on quantum resources.

Quantum information in gravitational background \cite{R1,R2,R3,R4,R5,R6,R7,R8,R9,BJ1,BJ2,BJ3,R10,R11,R12,R13,R14,R15,R16,R17,R18,RM1,RM2,RM3,RM4,RM5,RM6,RM7,RM8,RM9,RM10,RM11,RM12,QQRM12,WDD1,WDD2,WDD3,BJ4,WDD4,WDD5,B31,QQQQR8,qert1,RTYR4QQZ,qert2,qert3,qert4,qert5,qert6,qert7}, an interdisciplinary field that bridges quantum information theory, quantum field theory, and general relativity.
Theoretically, the influence of the gravitational effect on bipartite steering, entanglement, discord, coherence, quantum Fisher information, and entropic uncertainty relations has been investigated extensively in the background of a
dilaton black hole \cite{A1,A2,A3,A4,A5,A6,A7,A8,A9,A10,A11,A12,A13,A14,A15}.  As the complexity of quantum information tasks continues to increase, bipartite and tripartite inseparable states can no longer meet the growing demands, necessitating the use of N-partite inseparable states to handle these advanced tasks. For example,  Deng et al. have reported new Gaussian boson sampling experiments with pseudo-photon-number-resolving detection, which register up to 255 photon-click events \cite{Q13}, and Google quantum AI and collaborators have demonstrated surface code memory below threshold in new Willow architecture by using 105 nonseparable qubits \cite{Q14}. It is necessary to understand the N-partite nonseparable quantum systems in gravitational background, because the influence of gravitational effects on realistic quantum systems is a non-negligible factor.
However, measuring N-partite entanglement and obtaining its analytical expression in the dilaton black hole, particularly for complex W state, is a challenging task.
Thus, one of our primary motivations is to use the Abe-Rajagopal (AR) $q$-conditional entropy to quantify the nonseparability of N-partite systems under the dilaton black hole background.  Notably, this approach places stricter constraints on separability compared to traditional methods, such as the von Neumann conditional entropy, offering more nuanced insights into quantum correlations under the extreme gravitational background \cite{A17,A18,A19}.

Previous studies have indicated that fermionic coherence is generally weaker than bosonic coherence in the dilaton black hole background \cite{A11,A16}. However, the precise differences between the nonseparability of bosonic and fermionic states in dilaton spacetime remain largely unexplored. Similarly, while it has been shown that the coherence of the GHZ state is weaker than that of the W state in this context, the comparative nonseparability of these states within the specific framework of the GHS dilaton black hole has not been systematically investigated. Addressing these two critical questions provides another key motivation for our research.
Building on these motivations, we will study the impact of the Hawking effect on the nonseparability of N-partite GHZ and W states by using the AR $q$-conditional entropy for both free bosonic and fermionic fields in the background of the GHS dilaton black hole. Initially, we assume that $N$ observers are distributed in an asymptotically flat region, each sharing GHZ and W states. Next, we consider that $N-1$ observers remain stationary in the flat region, while a single observer is positioned near the event horizon of the black hole.

In this framework, our goal is to calculate the N-partite nonseparability for both bosonic and fermionic fields and derive the corresponding analytical expressions in curved spacetime. Our study yields four intriguing conclusions: (i) the Hawking effect from the black hole can create a net nonseparability in the W state for fermionic field; (ii) fermionic nonseparability is stronger than bosonic nonseparability, while fermionic coherence is weaker than bosonic coherence in the dilaton black hole \cite{A11,A16}; (iii) the GHZ state exhibits stronger nonseparability than the W state, yet its quantum coherence is weaker in dilaton spacetime; (iv) increasing the initial number of particles in the W state enhances quantum coherence in curved spacetime but does not favor nonseparability. These findings highlight the importance of selecting the appropriate quantum resources based on the type of particles involved, offering valuable insights for efficiently addressing relativistic quantum information challenges.

The paper is organized as follows. In Sec.II, we briefly outline the quantification of nonseparability using the AR $q$-conditional entropy in a quantum system.  In Sec.III, we discuss the quantization of both free bosonic and fermionic fields in the background of the GHS dilaton black hole.  In Sec.IV, we study bosonic and fermionic nonseparability of N-partite GHZ and W states  by employing the AR $q$-conditional entropy in the dilaton black hole.  Finally the summary is arranged in Sec.V.

\section{Abe-Rajagopal $q$-conditional entropy}
In this section, we briefly introduce the definition and measurement of AR $q$-conditional entropy \cite{A17,A18,A19}. Consider a system  divided into two subsystems, $A$ and $B$. Let $p_{ij}(A,B)$  represent the normalized joint probability of finding subsystem $A$ in its  $i$-th state and subsystem $B$ in its $j$-th state. Then the conditional probability of $B$ with $A$ found in its $i$-th state is given by $p_{ij}(B|A)=p_{ij}(A,B)/p_{i}(A)$, leading to the celebrated Bayes multiplication law
\begin{eqnarray}\label{S1}
p_{ij}(A,B)=p_{i}(A) p_{ij}(B|A),
\end{eqnarray}
where $p_{i}(A)$ represents the marginal probability distribution, defined as $p_{i}(A)=\sum_{j}p_{ij}(A,B)$.  From this law, we can derive the Boltzmann-Shannon entropy for the combined system of $A$ and $B$ as
\begin{eqnarray}\label{S2}
S(A,B)=-\sum_{i,j}p_{ij}(A,B)\ln p_{ij}(A,B).
\end{eqnarray}
Referred to as nonextensive statistical mechanics \cite{A20}, the Boltzmann-Shannon entropy in Eq.(\ref{S2}) can be generalized to
\begin{eqnarray}\label{S3}
S_{q}(A,B)=\frac{1}{1-q}\bigg\{\sum_{i,j}[p_{ij}(A,B)]^{q}-1\bigg\},
\end{eqnarray}
where $q$ is a positive parameter. In quantum theory, the probability distribution is replaced by the density matrix $\rho$, which is Hermitian, traceclass, and positive semidefinite, accommodating both pure and mixed states. When the state is pure, the equality $\hat{\rho}^{2}=\hat{\rho}$ holds, whereas $\hat{\rho}^{2}<\hat{\rho}$ for the mixed state. The quantum counterpart of Eq.(\ref{S3}) is given by
\begin{eqnarray}\label{S4}
S_{q}(A,B)=\frac{1}{1-q}\bigg[\mathrm{Tr}({\rho}_{AB})^{q}-1\bigg].
\end{eqnarray}
However, this nonadditivity condition is violated by the probability $p_{ij}(A,B)$ when subsystems $A$ and $B$ exhibit long-range interactions, such as entanglement, which persists even when they are spatially separated by a long distance.
To address this issue, a conditional form of the Tsallis $q$ entropy, introduced by Abe and Rajagopal, called the AR $q$-conditional entropy, can be defined as
\begin{eqnarray}\label{S5}
S_{q}(B|A)=\frac{S_{q}(A,B)-S_{q}(A)}{1+(1-q)S_{q}(A)}.
\end{eqnarray}
The computable form of this conditional entropy for any $A-B$ bipartition in a multipartite state can be obtained by substituting Eq.(\ref{S4}) into Eq.(\ref{S5})
\begin{eqnarray}\label{S6}\notag\
S_{q}(B|A)=\frac{1}{q-1}\Bigg[1-\frac{\mathrm{Tr}(\rho_{AB}^{q})}{\mathrm{Tr}(\rho_{A})^{q}}\Bigg]\\=\frac{1}{q-1}\Bigg[1-\frac{\Sigma_{n}\lambda_{n}^{q}(\rho_{AB})}{\Sigma_{m}\lambda_{m}^{q}(\rho_{A})}\Bigg],
\end{eqnarray}
where $\lambda_{n}$ and $\lambda_{m}$ represent the eigenvalues of the whole system $\rho_{AB}$ and one of its subsystem $\rho_{A}$, respectively. This conditional entropy adheres to the nonadditivity relation:$S_{q}(A,B)=S_{q}(A)+S_{q}(B)+(1-q)S_{q}(A)S_{q}(B)$. The AR $q$-conditional entropy yields negative values for nonseparable states, making the positivity of the AR $q$-conditional entropy a criterion for capturing separability in bipartite states. From this point forward, we quantify nonseparability using the AR $q$-conditional entropy for GHZ and W states, with one of the qubits positioned near the event horizon of the black hole.

\section{Quantization of  bosonic and fermionic fields in GHS dilaton spacetime}
The metric of a charged Garfinkle-Horowitz-Strominger (GHS) black hole spacetime becomes
\begin{eqnarray}\label{S7}
ds^{2}=-\bigg(\frac{r-2M}{r-2D}\bigg)dt^{2}+\bigg(\frac{r-2M}{r-2D}\bigg)^{-1}dr^{2}+r(r-2D)d\Omega^{2},
\end{eqnarray}
where $D$ and $M$ are parameters associated with the dilaton field and the mass of the black hole, respectively. The relationship among $D$, the magnetic charge $Q$, and $M$ is described by the equation $D=Q^{2}/2M$  \cite{rfv01,rfv02}. In this paper, we adopt natural units where $G=c=\hbar=\kappa_{B}=1$.

\subsection{Bosonic field}
The general perturbation equation for a coupled massive scalar field in this dilation spacetime is given by \cite{rfv2}
\begin{eqnarray}\label{S8}
\frac{1}{\sqrt{-g}}\partial_{\mu}(\sqrt{-g}g^{\mu\nu}\partial_{\nu})\Psi-(\mu+\xi R)\Psi=0,
\end{eqnarray}
where $\mu/\nu$ represents the coordinate components in spacetime, $\psi$ is the scalar field, and $R$ is the Ricci scalar curvature. The coupling between the gravitational field and the scalar field is represented by the term $\xi R \psi$, where $\xi$ is a numerical couple factor.
The normal mode solution can be expressed as
\begin{eqnarray}\label{S9}
\Psi_{\omega lm}=\frac{1}{h(r)}\chi_{\omega l}(r)Y_{lm}(\theta,\varphi)e^{-i\omega t},
\end{eqnarray}
where $Y_{lm}(\theta,\varphi)$ represents a scalar spherical harmonic on the unit two-sphere and $h(r)=\sqrt{r(r-2D)}$ with $D=Q^{2}/2M$. The radial equation can be obtained as
\begin{eqnarray}\label{S10}
\frac{d^{2}\chi_{\omega l}}{dr^{2}_{\ast}}+[\omega^{2}-V(r)]\chi_{\omega l}=0,
\end{eqnarray}
and
\begin{eqnarray}\label{S11}
V(r)=\frac{f(r)}{h(r)}\frac{d}{dr}\bigg[f(r)\frac{dh(r)}{dr}\bigg]+\frac{f(r)l(l+1)}{h^{2}(r)}+f(r)\bigg[\mu^{2}+\frac{2\xi D^{2}(r-2M)}{r^{2}(r-2D)^{3}}\bigg],
\end{eqnarray}
where $f(r)=(r-2M)/(r-2D)$ and the tortoise coordinate $r_{\ast}$ is defined  as $dr_{\ast}=dr/f(r)$ \cite{A5}.

To solve Eq.(\ref{S10}) near the event horizon, the resulting incoming wave function can be obtained, which is analytic everywhere in the spacetime manifold
\begin{eqnarray}\label{S12}
\Psi_{\mathrm{in},\omega lm}=e^{-i\omega v}Y_{lm}(\theta,\varphi),
\end{eqnarray}
and the outgoing wave functions for the regions outside and inside the event horizon are given by
\begin{eqnarray}\label{S13}
\Psi_{\mathrm{out},\omega lm}(r>r_{+})=e^{-i\omega u}Y_{lm}(\theta,\varphi),
\end{eqnarray}
\begin{eqnarray}\label{S14}
\Psi_{\mathrm{out},\omega lm}(r<r_{+})=e^{i\omega u}Y_{lm}(\theta,\varphi),
\end{eqnarray}
where $u=t-r_{\ast}$ and $v=t+r_{\ast}$. Eqs.(\ref{S13}) and (\ref{S14}) are individually analytic in the regions outside and inside the event horizon, respectively. As a result, they form a completely orthogonal family.
The generalized light-like Kruskal coordinates are defined as
\begin{eqnarray}\label{S15}
u&=&-4(M-D)\ln[-U/(4M-4D)],\notag\\v&=&4(M-D)\ln[V/(4M-4D)],\mathrm{if}\;r>r_{+},\notag\\
u&=&-4(M-D)\ln[U/(4M-4D)],\notag\\v&=&4(M-D)\ln[V/(4M-4D)],\mathrm{if}\;r<r_{+}.
\end{eqnarray}
Therefore, Eqs.(\ref{S13}) and (\ref{S14}) can be rewritten as
\begin{eqnarray}\label{S16}
\Phi_{\mathrm{out},\omega lm}(r>r_{+})=e^{4(M-D)i\omega\ln[U/(4M-4D)]}Y_{lm}(\theta,\varphi),
\end{eqnarray}
\begin{eqnarray}\label{S17}
\Phi_{\mathrm{out},\omega lm}(r<r_{+})=e^{-4(M-D)i\omega\ln[-U/(4M-4D)]}Y_{lm}(\theta,\varphi).
\end{eqnarray}
By utilizing the relation $-1=e^{i\pi}$ and ensuring that  Eq.(\ref{S16}) is analytic in the lower half-plane of $U$, a complete basis for positive-energy $U$ modes can be expressed as
\begin{eqnarray}\label{S18}
\Phi_{\mathrm{I},\omega lm}=e^{2\pi\omega(M-D)}\Phi_{\mathrm{out},\omega lm}(r>r_{+})
+e^{-2\pi\omega(M-D)}\Phi_{\mathrm{out},\omega lm}^{\ast}(r<r_{+}),
\end{eqnarray}
\begin{eqnarray}\label{S19}
\Phi_{\mathrm{II},\omega lm}=e^{-2\pi\omega(M-D)}\Phi_{\mathrm{out},\omega lm}^{\ast}(r>r_{+})
+e^{2\pi\omega(M-D)}\Phi_{\mathrm{out},\omega lm}(r<r_{+}).
\end{eqnarray}
Eqs.(\ref{S18}) and (\ref{S19}) are analytic for all real $U$ and $V$, which form a complete basis for positive-frequency modes.  Therefore, $\Phi_{\mathrm{I},\omega lm}$ and $\Phi_{\mathrm{II},\omega lm}$ can be employed to quantize the quantum field in Kruskal spacetime \cite{A6,A8}.

By applying second quantization to the field in the exterior region of the dilaton black hole, the Bogoliubov transformation for the creation and annihilation operators can be derived in both dilaton and Kruskal spacetimes
\begin{eqnarray}\label{S20}
a_{K,\omega lm}^{B,\dag}=\frac{1}{\sqrt{1-e^{-8\pi\omega(M-D)}}}b_{\mathrm{out},\omega lm}^{B,\dag}
-\frac{1}{\sqrt{e^{8\pi\omega(M-D)}-1}}b_{\mathrm{in},\omega lm}^{B},
\end{eqnarray}
\begin{eqnarray}\label{S21}
a_{K,\omega lm}^{B}=\frac{1}{\sqrt{1-e^{-8\pi\omega(M-D)}}}b_{\mathrm{out},\omega lm}^{B}-\frac{1}{\sqrt{e^{8\pi\omega(M-D)}-1}}b_{\mathrm{in},\omega lm}^{B,\dag},
\end{eqnarray}
where $a_{K,\omega lm}^{B,\dagger}$ and $a_{K,\omega lm}^{B}$ represent the bosonic creation and annihilation operators acting on the Kruskal vacuum of the exterior region, $b_{\mathrm{in},\omega lm}^{B,\dagger}$ and $b_{\mathrm{in},\omega lm}^{B}$ ($b_{\mathrm{out},\omega lm}^{B,\dagger}$ and $b_{\mathrm{out},\omega lm}^{B}$) represent the corresponding operators for the interior (exterior) region of the dilaton black hole. Therefore, the Kruskal vacuum $|0\rangle_{K}^{B}$ can be defined as $a_{K,\omega lm}^{B}|0\rangle_{K}^{B}=0$.
After normalizing the state vector appropriately, the Kruskal vacuum of the bosonic field in dilaton spacetime is a maximally entangled two-mode squeezed state
\begin{eqnarray}\label{S23}
|0\rangle_{K}^{B}=\sqrt{1-e^{-8\pi\omega(M-D)}}\sum^{\infty}_{n=0} e^{-4n\pi\omega(M-D)}|n\rangle_{\mathrm{out}}^{B}|n\rangle_{\mathrm{in}}^{B},
\end{eqnarray}
and the first excited state of the bosonic field reads
\begin{eqnarray}\label{S24}
|1\rangle_{K}^{B}=a_{K,\omega lm}^{\dagger}|0\rangle _{K}^{B}=
\big[1-e^{-8\pi\omega(M-D)}\big]\sum^{\infty}_{n=0}\sqrt{n+1} e^{-4n\pi\omega(M-D)}|n+1\rangle_{\mathrm{out}}^{B}|n\rangle_{\mathrm{in}}^{B},
\end{eqnarray}
where $B$ represents the bosonic field, $\{|n\rangle_{\mathrm{out}}\}$ and $\{|n\rangle_{\mathrm{in}}\}$ are the orthonormal bases for the outside and inside regions of the event horizon, respectively \cite{A7}.
For an external observer situated outside the dilaton black hole, it is imperative to trace over the modes within the interior region. This necessity arises because these modes are situated in a causally disconnected region, rendering them inaccessible to the observer. Consequently, the Hawking radiation spectrum for the bosonic field is derived as
\begin{eqnarray}\label{S25}
N^{B}_{\omega}=\frac{1}{e^{8\pi\omega(M-D)}-1}.
\end{eqnarray}
Eq.(\ref{S25}) reveals that an observer positioned outside the GHS dilaton black hole perceives a thermal Bose-Einstein distribution of particles while traversing the Kruskal vacuum.

\subsection{Fermionic field}
Similar to the bosonic field, the Bogoliubov transformations relating the Kruskal operators to the dilaton operators for the fermionic field can be expressed as
\begin{eqnarray}\label{S26}
a^{F,\dag}_{K,\omega lm}=\frac{1}{\sqrt{e^{-8\pi\omega(M-D)}+1}}a^{F,\dag}_{\mathrm{out},\omega lm}-\frac{1}{\sqrt{e^{8\pi\omega(M-D)}+1}}b_{\mathrm{in},\omega lm}^{F},
\end{eqnarray}
\begin{eqnarray}\label{S27}
a^{F}_{K,\omega lm}=\frac{1}{\sqrt{e^{-8\pi\omega(M-D)}+1}}a_{\mathrm{out},\omega lm}^{F}-\frac{1}{\sqrt{e^{8\pi\omega(M-D)}+1}}b^{F,\dag}_{\mathrm{in},\omega lm}.
\end{eqnarray}
Consequently, the Kruskal vacuum state and the excited states of the fermionic field in dilaton spacetime can be written as
\begin{eqnarray}\label{S28}
|0\rangle_{K}^{F}=\frac{1}{\sqrt{e^{-8\pi\omega(M-D)}+1}}|0\rangle_{\mathrm{out}}^{F}|0\rangle_{\mathrm{in}}^{F}+\frac{1}{\sqrt{e^{8\pi\omega(M-D)}+1}}|1\rangle_{\mathrm{out}}^{F}|1\rangle_{\mathrm{in}}^{F},
\end{eqnarray}
and
\begin{eqnarray}\label{S29}
|1\rangle_{K}^{F}=|1\rangle_{\mathrm{out}}^{F}|0\rangle_{\mathrm{in}}^{F},
\end{eqnarray}
where $F$ represents the fermionic field \cite{A10,A12}. The Hawking radiation spectrum of the fermionic field can be derived as
\begin{eqnarray}\label{S30}
N^{F}_{\omega}=\frac{1}{e^{8\pi\omega(M-D)}+1}.
\end{eqnarray}
This result indicates that an observer outside the event horizon detects a thermal Fermi-Dirac distribution of particles \cite{A1}.
From Eqs. (\ref{S25}) and (\ref{S30}), it becomes evident that the Bose-Einstein and Fermi-Dirac distributions governing bosons and fermions respectively generate fundamentally distinct gravitational responses in dilaton spacetime. These divergence effects manifest through characteristically different modulations of coherence and entanglement in bosonic versus fermionic fields, ultimately underscoring how quantum statistical signatures of the two systems diverge markedly in the vicinity of a dilaton black hole.

\section{N-partite nonseparability of bosonic and fermionic fields in dilaton spacetime}

In this section, we first examine multipartite GHZ and W states in the context of a particle located near the black hole event horizon. We then characterize their nonseparability in the nongravitational-gravitational bipartition using the AR $q$-conditional entropy, as defined in Eq.(\ref{S6}), with respect to the parameter $q$ and the dilaton $D$ in curved spacetime.
Initially, we assume that  $N(N\geq3)$ observers share a GHZ state
\begin{eqnarray}\label{S31}
|\mathrm{GHZ}_{123\ldots N}^{B/F}\rangle=\frac{1}{\sqrt{2}}\left(|00\ldots00\rangle+|11\ldots11\rangle\right),
\end{eqnarray}
and a W state
\begin{eqnarray}\label{S32}
|\mathrm{W}_{123\ldots N}^{B/F}\rangle=\frac{1}{\sqrt{N}}\left(|10\ldots00\rangle+|01\ldots00\rangle+\ldots+|00\ldots01\rangle\right),
\end{eqnarray}
where the mode $i(i=1,2,\ldots,N)$ is detected by observer $O_{i}$, at the same point in the asymptotically flat region of the dilaton black hole. Here, we use $B$ and $F$ to denote bosonic and fermionic modes, respectively.  After sharing their qubits, one observer hovers near the event horizon of the dilaton black hole, while the remaining $N-1$ observers remain stationary in the asymptotically flat region.

\subsection{N-partite nonseparability of bosonic field}
By using the transformations from Kruskal modes to dilaton modes as given in Eqs.(\ref{S23}) and (\ref{S24}), N-partite GHZ state of Eq.(\ref{S31}) can be rewritten as
\begin{eqnarray}\label{S33}
|\mathrm{GHZ}^{B}_{123\ldots N+1}\rangle &=&\frac{1}{\sqrt{2}}\bigg\{[1-e^{-8\pi\omega(M-D)}]^{\frac{1}{2}}\overbrace{(|0\rangle_{1}|0\rangle_{2}\cdots|0\rangle_{N-1})}^{|\overline{0}\rangle}\sum^{\infty}_{n=0}\big[e^{-4n\pi\omega(M-D)}\notag\\
&&|n\rangle_{N,\mathrm{out}}|n\rangle_{N+1,\mathrm{in}}\big]+[1-e^{-8\pi\omega(M-D)}]\overbrace{(|1\rangle_{1}|1\rangle_{2}\cdots|1\rangle_{N-1})}^{|\overline{1}\rangle}\notag\\
&&\sum^{\infty}_{m=0}\big[e^{-4m\pi\omega(M-D)}\sqrt{m+1}|m+1\rangle_{N,\mathrm{out}}|m\rangle_{N+1,\mathrm{in}}\big]\bigg\},
\end{eqnarray}
where $|\overline{0}\rangle=|0\rangle_{1}|0\rangle_{2}\cdots|0\rangle_{N-1}$ and $|\overline{1}\rangle=|1\rangle_{1}|1\rangle_{2}\cdots|1\rangle_{N-1}$.
Since the region outside the event horizon is causally disconnected from the interior, we should trace out the modes within the event horizon. After tracing out these inaccessible modes, we obtain the mixed density matrix for the exterior region
\begin{eqnarray}\label{S34}
\rho_{123\ldots N_{\mathrm{out}}}^{B,\mathrm{GHZ}}
&=&\frac{1}{2}\sum_{n}^{\infty}e^{-8n\pi\omega(M-D)}\bigg\{[1-e^{-8\pi\omega(M-D)}]|\overline{0}\rangle\langle\overline{0}||n\rangle_{N,\mathrm{out}}\langle n|\notag\\
&+&[1-e^{-8\pi\omega(M-D)}]^{\frac{3}{2}}\sqrt{n+1}|\overline{0}\rangle\langle\overline{1}||n\rangle_{N,\mathrm{out}}\langle n+1|\notag\\
&+&[1-e^{-8\pi\omega(M-D)}]^{\frac{3}{2}}\sqrt{n+1}|\overline{1}\rangle\langle\overline{0}|| n+1\rangle_{N,\mathrm{out}}\langle n|\notag\\
&+&[1-e^{-8\pi\omega(M-D)}]^{2}(n+1)|\overline{1}\rangle\langle\overline{1}|| n+1\rangle_{N,\mathrm{out}}\langle n+1|\bigg\}.
\end{eqnarray}
The structure of this density matrix consists of  $2\times2$ blocks along the diagonal, with all off-diagonal elements equal to zero. It is represented by
\begin{eqnarray}\label{S35}
\rho_{123\ldots N_{\mathrm{out}}}^{B,\mathrm{GHZ}}=\frac{1}{2}\begin{pmatrix}
0 & & & & & \\
& \Delta_0 & & & & \\
& & \Delta_1 & & & \\
& & & \ddots & & \\
& & & & \Delta_n & \\
& & & & & \ddots \\
\end{pmatrix},
\end{eqnarray}
where
\begin{eqnarray}\label{S36}
\Delta_{n}(\rho_{123\ldots N_{\mathrm{out}}}^{B,\mathrm{GHZ}})=\begin{pmatrix}
(1-\alpha)\alpha^{n} & (1-\alpha)^{\frac{3}{2}}\sqrt{n+1}\alpha^{n} \\
(1-\alpha)^{\frac{3}{2}}\sqrt{n+1}\alpha^{n} & (1-\alpha)^{2}(n+1)\alpha^{n}  \\
\end{pmatrix},
\end{eqnarray}
with $\alpha=e^{-8\pi\omega(M-D)}$. The eigenvalues of this n-th block of the state $\rho_{123\ldots N_{\mathrm{out}}}^{B,\mathrm{GHZ}}$ are 0 and
\begin{eqnarray}\label{S37}
\lambda_{\mathrm{GHZ}_{n}}^{B}=\frac{1}{2}[1-e^{-8\pi\omega(M-D)}]e^{-8n\pi\omega(M-D)}[2+n-(n+1)e^{-8\pi\omega(M-D)}].
\end{eqnarray}
Therefore, the trace of the density matrix satisfies $\mathrm{Tr}[\rho_{123\ldots N_{\mathrm{out}}}^{B,\mathrm{GHZ}}]=\sum_{n=0}^{\infty}\lambda_{\mathrm{GHZ}_{n}}^{B}=1$. \par
The reduced density matrix $\rho_{N_{\mathrm{out}}}^{B,\mathrm{GHZ}}$, corresponding to the modes of the last qubit in exterior region of the dilaton black hole, is obtained by tracing out subsystems of the front $N-1$ particles, i.e., $\rho_{N_{\mathrm{out}}}^{B,\mathrm{GHZ}}=\mathrm{Tr}_{123\ldots N-1}[\rho_{123\ldots N_{\mathrm{out}}}^{B,\mathrm{GHZ}}]$, and is given by
\begin{eqnarray}\label{S38}
\rho_{N_{\mathrm{out}}}^{B,\mathrm{GHZ}}&=&\frac{1}{2}\big[1-e^{-8\pi\omega(M-D)}\big]\sum_{m=0}^{\infty}e^{-8m\pi\omega(M-D)}\bigg\{1+m\notag\\
&\times&\big[1-e^{-8\pi\omega(M-D)}\big]e^{8\pi\omega(M-D)}\bigg\} |m\rangle_{N,\mathrm{out}}\langle m|.
\end{eqnarray}
This is an infinite-dimensional diagonal matrix whose $m$-th eigenvalue is given by
\begin{eqnarray}\label{S39}
\lambda_{\mathrm{GHZ}_{m}}^{B}=\frac{1}{2}[1-e^{-8\pi\omega(M-D)}]e^{-8m\pi\omega(M-D)}\bigg\{1+m[1-e^{-8\pi\omega(M-D)}]e^{8\pi\omega(M-D)}\bigg\}.
\end{eqnarray}
Here $\mathrm{Tr}[\rho_{N_{\mathrm{out}}}^{B,\mathrm{GHZ}}]=\sum_{m=0}^{\infty}\lambda_{\mathrm{GHZ}_{m}}^{B}=1$. \par
To characterize the nonseparability of the state $|\mathrm{GHZ}^{B}_{123\ldots N+1}\rangle $  in dilaton spacetime, we substitute the eigenvalues $\lambda_{\mathrm{GHZ}_{n}}^{B}$ of $\rho_{123\ldots N_{\mathrm{out}}}^{B,\mathrm{GHZ}}$ [see Eq.(\ref{S37})] and $\lambda_{\mathrm{GHZ}_{m}}^{B}$ of $\rho_{N_{\mathrm{out}}}^{B,\mathrm{GHZ}}$ [see Eq.(\ref{S39})] in the AR $q$- conditional entropy given in Eq.(\ref{S6}) to obtain
\begin{eqnarray}\label{S40}
S_{B,q}^{\mathrm{GHZ},N}=\frac{1}{q-1}\bigg\{1-\frac{\sum_{n=0}^{\infty}\big\{e^{-8n\pi\omega(M-D)}\big[2+n-(n+1)e^{-8\pi\omega(M-D)}\big]\big\}^{q}}{\sum_{m=0}^{\infty}\big\{e^{-8m\pi\omega(M-D)}\big[1+m[1-e^{-8\pi\omega(M-D)}]e^{8\pi\omega(M-D)}\big]\big\}^{q}}\bigg\}.
\end{eqnarray}
The negative values of this AR $q$-conditional entropy $S_{B,q}^{\mathrm{GHZ},N}$ characterize the nonseparability between nongravitational-gravitational bipartition outside the event horizon of the GHS dilaton black hole. From Eq.(\ref{S40}), we can see that the Hawking effect from the black hole can influence  the bosonic nonseparability of the GHZ state. Importantly, increasing the number of qubits in the asymptotically flat region does not affect the nonseparability behavior of the  N-qubit GHZ state when one of its qubits is dilated.

Similarly, the pure  N-partite W state can be rewritten in terms of dilaton modes as
\begin{eqnarray}\label{S41}
|\mathrm{W}^{B}_{123\ldots N+1}\rangle
&=&\frac{1}{\sqrt{N}}\bigg\{[1-e^{-8\pi\omega(M-D)}]^{\frac{1}{2}}|1\rangle_{_{1}}|0\rangle_{_{2}}\cdots|0\rangle_{_{N-1}}\sum^{\infty}_{n=0}\big[e^{-4n\pi\omega(M-D)}|n\rangle_{N,\mathrm{out}}|n\rangle_{N+1,\mathrm{in}}\big]\notag\\
&+&[1-e^{-8\pi\omega(M-D)}]^{\frac{1}{2}}|0\rangle_{_{1}}|1\rangle_{_{2}}\cdots|0\rangle_{_{N-1}}\sum^{\infty}_{n=0}\big[e^{-4n\pi\omega(M-D)}|n\rangle_{N,\mathrm{out}}|n\rangle_{N+1,\mathrm{in}}\big]\notag\\
&+&\cdots+[1-e^{-8\pi\omega(M-D)}]^{\frac{1}{2}}|0\rangle_{_{1}}|0\rangle_{_{2}}\cdots|1\rangle_{_{N-1}}\sum^{\infty}_{n=0} \big[ e^{-4n\pi\omega(M-D)}|n\rangle_{N,\mathrm{out}}|n\rangle_{N+1,\mathrm{in}}\big]\notag\\
&+&[1-e^{-8\pi\omega(M-D)}]|\overline{0}\rangle\sum^{\infty}_{n=0}\big[\sqrt{n+1}e^{-4n\pi\omega(M-D)}|n+1\rangle_{N,\mathrm{out}}|n\rangle_{N+1,\mathrm{in}}\big]\bigg\}.
\end{eqnarray}
After tracing out the modes inside the event horizon, we obtain the density operator $\rho_{123\ldots N_{\mathrm{out}}}^{B,\mathrm{W}}$   as
\begin{eqnarray}\label{S42}
\rho_{123\ldots N_{\mathrm{out}}}^{B,\mathrm{W}}=\frac{1}{N}\big(\rho_{\mathrm{diag}}^{B}+\rho_{\mathrm{off}}^{B}\big),
\end{eqnarray}
where the diagonal part $\rho_{\mathrm{diag}}^{B}$ is given by
\[
\begin{aligned}
\rho_{\mathrm{diag}}^{B} &= \sum_{i=1}^{N-1} |1\rangle_{i} \langle 1| \bigotimes_{j=1, j \neq i}^{N-1} |0\rangle_{j} \langle 0| [1 - e^{-8\pi \omega (M - D)}] \sum_{n=0}^{\infty} e^{-8n\pi \omega (M - D)} |n\rangle_{N, \mathrm{out}} \langle n| \\
&\quad + \bigotimes_{i=1}^{N-1} |0\rangle_{i} \langle 0| [1 - e^{-8\pi \omega (M - D)}]^2 \sum_{n=0}^{\infty} (n+1) e^{-8n\pi \omega (M - D)} |n+1\rangle_{N, \mathrm{out}} \langle n+1|,
\end{aligned}
\]
and the off-diagonal part $\rho_{\mathrm{off}}^{B}$ reads
\[
\begin{aligned}
\rho_{\mathrm{off}}^{B} &=\sum_{i,j=1(i\neq j)}^{N-1}|0\rangle_{i}\langle1||1\rangle_{j}\langle0|\bigotimes_{k=1(k\neq i,k\neq j)}^{N-1}|0\rangle_{k}\langle0|[1-e^{-8\pi\omega(M-D)}]\sum_{n=0}^{\infty}e^{-8n\pi\omega(M-D)}|n\rangle_{N,\mathrm{out}}\langle n| \\
&+\sum_{i=1}^{N-1}|1\rangle_{i}\langle0|\bigotimes_{j=1(j\neq i)}^{N-1}|0\rangle_{j}\langle0|[1-e^{-8\pi\omega(M-D)}]^{\frac{3}{2}}\sum_{n=0}^{\infty}\sqrt{n+1}e^{-8n\pi\omega(M-D)}|n\rangle_{N,\mathrm{out}}\langle n+1| \\
&+\sum_{i=1}^{N-1}|0\rangle_{i}\langle1|\bigotimes_{j=1(j\neq i)}^{N-1}|0\rangle_{j}\langle0|[1-e^{-8\pi\omega(M-D)}]^{\frac{3}{2}}\sum_{n=0}^{\infty}\sqrt{n+1}e^{-8n\pi\omega(M-D)}|n+1\rangle_{N,\mathrm{out}}\langle n|.
\end{aligned}
\]
The density matrix of this state $\rho_{123\ldots N_{\mathrm{out}}}^{B,\mathrm{W}}$ assumes a block structure along the diagonal with all off-diagonal elements being zero and can be expressed as
\begin{eqnarray}\label{S43}
\rho_{123\ldots N_{\mathrm{out}}}^{B,\mathrm{W}}=\frac{1}{N}\begin{pmatrix}
0 & & & & & \\
& \Delta'_0 & & & & \\
& & \Delta'_1 & & & \\
& & & \ddots & & \\
& & & & \Delta'_n & \\
& & & & & \ddots \\
\end{pmatrix},
\end{eqnarray}
where the diagonal blocks $\Delta'_n$ are represented by
\begin{eqnarray}\label{S44}
&&\Delta'_{n}(\rho_{123\ldots N_{\mathrm{out}}}^{B,W})=\notag\\
&&\alpha^{n}\begin{pmatrix}
(1-\alpha) & (1-\alpha) & \cdots & (1-\alpha) & (1-\alpha)^\frac{3}{2}\sqrt{n+1}\\
(1-\alpha) & (1-\alpha) & \cdots & (1-\alpha) & (1-\alpha)^\frac{3}{2}\sqrt{n+1}\\
\vdots & \vdots & \ddots &\vdots & \vdots \\
(1-\alpha) & (1-\alpha) & \cdots & (1-\alpha) & (1-\alpha)^\frac{3}{2}\sqrt{n+1}\\
(1-\alpha)^\frac{3}{2}\sqrt{n+1} & (1-\alpha)^\frac{3}{2}\sqrt{n+1} & \cdots & (1-\alpha)^\frac{3}{2}\sqrt{n+1} & (1-\alpha)^{2}(n+1)
\end{pmatrix},
\end{eqnarray}
which is a block matrix in the subspace of $\{|10\ldots0n\rangle, |01\ldots0n\rangle, \cdots |00\ldots1n\rangle, |00\ldots0n+1\rangle \} $.
The matrix $\Delta'_{n}(\rho_{123\ldots N_{\mathrm{out}}}^{B,\mathrm{W}})$ has a specific block structure. It is a $N\times N$ matrix, where the upper-left $(N-1)\times(N-1)$ submatrix consists entirely of the same element, $(1-\alpha)\alpha^n$. The entries in both the last row and the last column are all equal to $(1-\alpha)^{\frac{3}{2}}\sqrt{n+1}\alpha^{n}$, with the exception of the bottom-right corner element, which is $(1-\alpha)^{2}(n+1)\alpha^n$.

The eigenvalues of this $n$-th block of the state $\rho_{123\ldots N_{out}}^{B,\mathrm{W}}$ are 0 and
\begin{eqnarray}\label{S45}
\lambda^B_{W_{n}}=\frac{1}{N}e^{-8n\pi\omega(M-D)}[1-e^{-8\pi\omega(M-D)}][N+n-(n+1)e^{-8\pi\omega(M-D)}].
\end{eqnarray}
The reduced density matrix $\rho_{N_{\mathrm{out}}}^{B,\mathrm{W}}$ is obtained by tracing out $N-1$ subsystems as
\begin{eqnarray}\label{S46}
\rho_{N_{\mathrm{out}}}^{B,\mathrm{W}} &=& \frac{1}{N}\sum_{m=0}^{\infty}e^{-8m\pi\omega(M-D)}[1-e^{-8\pi\omega(M-D)}]\bigg\{[1-e^{-8\pi\omega(M-D)}](m+1)|m+1\rangle_{N,\mathrm{out}}\langle m+1|\notag\\
&+& (N-1)|m\rangle_{N,\mathrm{out}}\langle m|\bigg\}=\frac{1}{N}e^{-8m\pi\omega(M-D)}[1-e^{-8\pi\omega(M-D)}]\bigg\{N-1+me^{8\pi\omega(M-D)}\notag\\
&\times&[1-e^{-8\pi\omega(M-D)}]\bigg\}|m\rangle_{N,\mathrm{out}}\langle m|,
\end{eqnarray}
and the $m$-th eigenvalue of diagonal matrix above is computed as
\begin{eqnarray}\label{S47}
\lambda^B_{\mathrm{W}_{m}}=\frac{1}{N}e^{-8m\pi\omega(M-D)}[1-e^{-8\pi\omega(M-D)}]\bigg\{N-1+me^{8\pi\omega(M-D)}[1-e^{-8\pi\omega(M-D)}]\bigg\}.
\end{eqnarray}
Substituting the eigenvalues $\lambda_{\mathrm{W}_{n}}^{B}$ of $\rho_{123\ldots N_{\mathrm{out}}}^{B,\mathrm{W}}$ [see Eq.(\ref{S45})] and $\lambda_{\mathrm{GHZ}_{m}}^{B}$ of $\rho_{N_{\mathrm{out}}}^{B,\mathrm{GHZ}}$ [see Eq.(\ref{S47})] into the AR $q$-conditional entropy given in Eq.(\ref{S6}), we obtain
 \begin{eqnarray}\label{S48}
S_{B,q}^{\mathrm{W},N}=\frac{1}{q-1}\left\{1-\frac{\sum_{n=0}^{\infty}\big\{[e^{-8n\pi\omega(M-D)}[N+n-(n+1)e^{-8\pi\omega(M-D)}]\big\}^{q}}{\sum_{m=0}^{\infty}\big\{e^{-8m\pi\omega(M-D)}\big[N-1+me^{8\pi\omega(M-D)}[1-e^{-8\pi\omega(M-D)}]\big]\big\}^{q}}\right\}.
\end{eqnarray}
From Eqs.(\ref{S40}) and (\ref{S48}), we observe that the nonseparability of the W state depends on the initial particle number $N$, whereas the nonseparability of the GHZ state is independent of  $N$ in dilaton spacetime. This distinction highlights the differing entanglement properties of these states under the influence of the Hawking effect of the black hole.

\begin{figure}
\begin{minipage}[t]{0.5\linewidth}
\centering
\includegraphics[width=3.0in,height=5.2cm]{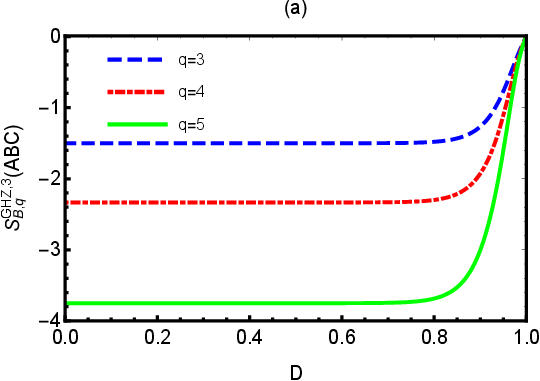}
\label{fig1a}
\end{minipage}%
\begin{minipage}[t]{0.5\linewidth}
\centering
\includegraphics[width=3.0in,height=5.2cm]{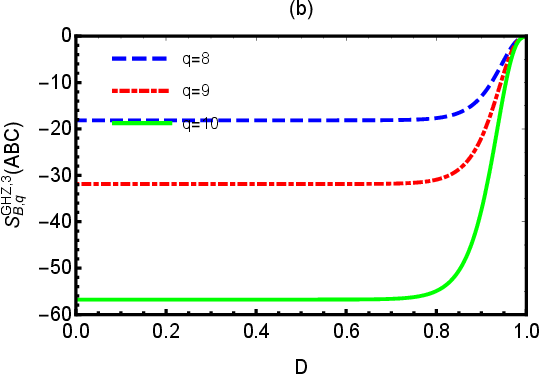}
\label{fig1b}
\end{minipage}%

\begin{minipage}[t]{0.5\linewidth}
\centering
\includegraphics[width=3.0in,height=5.2cm]{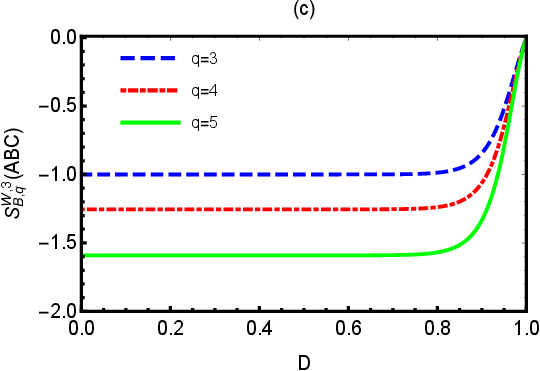}
\label{fig1c}
\end{minipage}%
\begin{minipage}[t]{0.5\linewidth}
\centering
\includegraphics[width=3.0in,height=5.2cm]{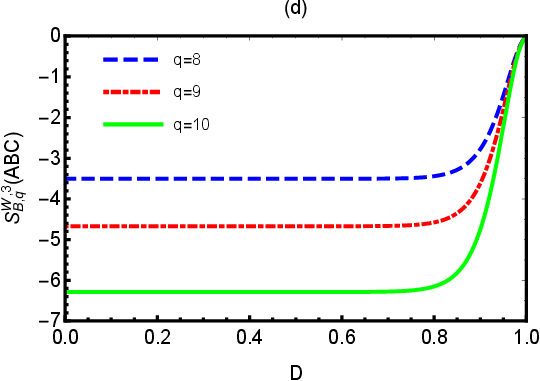}
\label{fig1d}
\end{minipage}%
\caption{AR $q$-conditional entropy for GHZ and W states of bosonic field as a function of the dilaton $D$ for various values of $q$ with $M=\omega=1$ and $N=3$.}
\label{Fig.1}
\end{figure}

In Fig.\ref{Fig.1}, we plot AR $q$-conditional entropy $S^{\mathrm{GHZ},_{3}}_{B,_{q}}(ABC)$ and $S^{\mathrm{W},_{3}}_{B,_{q}}(ABC)$ as a function of the dilaton $D$ for different $q$, derived from  Eqs.(\ref{S40}) and (\ref{S48}).  From Fig.\ref{Fig.1}, we observe that $q$-conditional entropy remains negative and increases monotonically with the increase of the dilaton $D$, where more negative values indicate stronger nonseparability.  This implies that the nonseparability of both GHZ and W states decreases  as the
dilaton $D$ increases. Notably, as the dilaton  $D$ approaches the mass  $M$ of the black hole, $q$-conditional entropy $S^{\mathrm{GHZ},_{3}}_{B,_{q}}(ABC)$ and $S^{\mathrm{W},_{3}}_{B,_{q}}(ABC)$
tend toward zero for all values of any $q$, signaling a transition from nonseparability to separability in the extreme limit $D\rightarrow M$. This marks a significant shift in the entanglement properties of quantum states at this threshold. Furthermore, the $q$-conditional entropy of W state is consistently larger than that of GHZ state within dilaton spacetime, indicating weaker nonseparability for W state under the influence of the Hawking effect. In contrast, quantum coherence of W state surpasses that of GHZ state in the dilaton black hole \cite{A16}. This contrast underscores the distinct behaviors of GHZ and W states in relativistic quantum information tasks.
In strong gravitational backgrounds (such as black hole), the Hawking effect can induce decoherence, making the stability of quantum systems crucial during task execution. Although GHZ state provides stronger multipartite entanglement and is suitable for tasks that require high entanglement (such as quantum teleportation and quantum key distribution), their comparatively weaker coherence makes it more susceptible to decoherence in strong gravitational fields. In contrast, W state, despite possessing weaker entanglement, exhibits stronger coherence and is therefore more resistant to gravity-induced decoherence. As a result, in practical relativistic quantum information tasks, it is essential to choose appropriate quantum resources based on specific task requirements: GHZ state is preferable for high-entanglement tasks, while W state is better suited for scenarios demanding stronger noise resistance. This finding highlights the importance of carefully balancing and selecting quantum resources in curved spacetime to enhance the performance and robustness of quantum protocols.

\subsection{N-partite nonseparability of fermionic field}
Similar to the GHZ state of bosonic field, we use the dilaton modes defined in Eqs.(\ref{S28}) and (\ref{S29}) to rewrite Eq.(\ref{S31}) of fermionic field as
\begin{eqnarray}\label{S49}
&&|\mathrm{GHZ}^{F}_{123\ldots N+1}\rangle=\frac{1}{\sqrt{2}}\bigg\{\overbrace{\big(|0\rangle_{1}|0\rangle_{2}\cdots|0\rangle_{N-1}\big)}^{|\bar{0}\rangle}\{[e^{-8\pi\omega(M-D)}+1]^{-\frac{1}{2}}|0\rangle_{N,\mathrm{out}}|0\rangle_{N+1,\mathrm{in}}\notag\\
&&+[e^{8\pi\omega(M-D)}+1]^{-\frac{1}{2}}|1\rangle_{N,\mathrm{out}}|1\rangle_{N+1,\mathrm{in}}\}+\overbrace{\big(|1\rangle_{1}|1\rangle_{2}\cdots|1\rangle_{N-1}\big)}^{|\bar{1}\rangle}|1\rangle_{N,\mathrm{out}}|0\rangle_{N+1,\mathrm{in}}\bigg\}.
\end{eqnarray}
After tracing over the inaccessible modes inside the event horizon, we obtain the density operator $\rho_{123\ldots N_{\mathrm{out}}}^{F,\mathrm{GHZ}}$ as
\begin{eqnarray}\label{S50}
\rho^{F,\mathrm{GHZ}}_{123\ldots N_{\mathrm{out}}}&=&\frac{1}{2}\bigg\{|\overline{0}\rangle\langle\overline{0}|\big\{[e^{-8\pi\omega(M-D)}+1]^{-1}|0\rangle_{N,\mathrm{out}}\langle0|+[e^{8\pi\omega(M-D)}+1]^{-1}|1\rangle_{N,\mathrm{out}}\langle1|\big\}\notag\\
&+&|\overline{0}\rangle\langle\overline{1}|[e^{-8\pi\omega(M-D)}+1]^{-\frac{1}{2}}|0\rangle_{N,\mathrm{out}}\langle1|+|\overline{1}\rangle\langle\overline{0}|[e^{-8\pi\omega(M-D)}+1]^{-\frac{1}{2}}|1\rangle_{N,\mathrm{out}}\langle0|\notag\\
&+&|\overline{1}\rangle\langle\overline{1}||1\rangle_{N,\mathrm{out}}\langle1|\bigg\}.
\end{eqnarray}
The eigenvalues of this state can be computed as
\begin{eqnarray}\label{S51}
\lambda^{F}_{\mathrm{GHZ}_{1}}=\frac{1}{2[1+e^{8\pi\omega(M-D)}]}, \quad \lambda^{F}_{\mathrm{GHZ}_{2}}=\frac{1+2e^{8\pi\omega(M-D)}}{2[1+e^{8\pi\omega(M-D)}]}.
\end{eqnarray}
The reduced density matrix $\rho_{N_{\mathrm{out}}}^{F,\mathrm{GHZ}}$ is obtained by tracing out subsystems of the front $N-1$ particles, and is given by
\begin{eqnarray}\label{S52}
\rho_{N_{\mathrm{out}}}^{F,\mathrm{GHZ}}=\frac{1}{2}\bigg\{[1+e^{-8\pi\omega(M-D)}]|0\rangle_{N,\mathrm{out}}\langle0|+\big\{[e^{8\pi\omega(M-D)}+1]^{-1}+1\big\}|1\rangle_{N,\mathrm{out}}\langle1|\bigg\},
\end{eqnarray}
with eigenvalues:
\begin{eqnarray}\label{S53}
\lambda^{F1}_{\mathrm{GHZ}_{1}}=\frac{e^{8\pi\omega(M-D)}}{2[1+e^{8\pi\omega(M-D)}]}, \quad \lambda^{F1}_{\mathrm{GHZ}_{2}}=\frac{2+e^{8\pi\omega(M-D)}}{2[1+e^{8\pi\omega(M-D)}]}.
\end{eqnarray}
Substituting these eigenvalues into Eq.(\ref{S6}), we obtain
\begin{eqnarray}\label{S54}
S_{F,q}^{\mathrm{GHZ},N}=\frac{1}{q-1}\bigg\{1-\frac{1+[1+2e^{8\pi\omega(M-D)}]^q}{e^{8\pi\omega(M-D)q}+[2+e^{8\pi\omega(M-D)}]^q}\bigg\}.
\end{eqnarray}
Similarly, the pure  $N$-partite W state [see Eq.(\ref{S32})] for fermionic field, re-expressed in terms of dilaton modes [Eqs.(\ref{S28}) and (\ref{S29})], is
\begin{eqnarray}\label{S55}
|\mathrm{W}^{F}_{123\ldots N+1}\rangle
&=&\frac{1}{\sqrt{N}}\bigg\{|1\rangle_{_{1}}|0\rangle_{_{2}}\cdots|0\rangle_{_{N-1}}\{[e^{-8\pi\omega(M-D)}+1]^{-\frac{1}{2}}|0\rangle_{N,\mathrm{out}}|0\rangle_{N+1,\mathrm{in}}\notag\\
&+&(e^{8\pi\omega(M-D)}+1)^{-\frac{1}{2}}|1\rangle_{N,\mathrm{out}}|1\rangle_{N+1,\mathrm{in}}\}+|0\rangle_{_{1}}|1\rangle_{_{2}}\cdots|0\rangle_{_{N-1}}\big\{[e^{-8\pi\omega(M-D)}+1]^{-\frac{1}{2}}\notag\\
&\times&|0\rangle_{N,\mathrm{out}}|0\rangle_{N+1,\mathrm{in}}+[e^{8\pi\omega(M-D)}+1]^{-\frac{1}{2}}|1\rangle_{N,\mathrm{out}}|1\rangle_{N+1,\mathrm{in}}\big\}+\cdots\notag\\
&+&|0\rangle_{_{1}}|0\rangle_{_{2}}\cdots|1\rangle_{_{N-1}}\big\{[e^{-8\pi\omega(M-D)}+1]^{-\frac{1}{2}}|0\rangle_{N,\mathrm{out}}|0\rangle_{N+1,\mathrm{in}}+[e^{8\pi\omega(M-D)}+1]^{-\frac{1}{2}}\notag\\
&\times&|1\rangle_{N,\mathrm{out}}|1\rangle_{N+1,\mathrm{in}}\big\}+|0\rangle_{1}|0\rangle_{2}\cdots|0\rangle_{N-1}|1\rangle_{N,\mathrm{out}}|0\rangle_{N+1,\mathrm{in}}\bigg\}.
\end{eqnarray}
Tracing out the modes inside the event horizon, the density operator $\rho_{123\ldots N_{\mathrm{out}}}^{F,\mathrm{W}}$  reads
\begin{eqnarray}\label{S56}
\rho_{123\ldots N_{\mathrm{out}}}^{F,\mathrm{W}}=\frac{1}{N}\big(\rho_{\mathrm{diag}}^{F}+\rho_{\mathrm{off}}^{F}\big),
\end{eqnarray}
where the diagonal part $\rho_{\mathrm{diag}}^{F}$ is given by
\[
\begin{aligned}
\rho_{\mathrm{diag}}^{F}&=\sum_{i=1}^{N-1}|1\rangle_{i}\langle1|\bigotimes_{j=1(j\neq i)}^{N-1}|0\rangle_{j}\langle0|\big\{[e^{-8\pi\omega(M-D)}+1]^{-1}|0\rangle_{N,\mathrm{out}}\langle0| \\
&+[e^{8\pi\omega(M-D)}+1]^{-1}|1\rangle_{N,\mathrm{out}}\langle1|\big\}+\bigotimes_{i=1}^{N-1}|0\rangle_{i}\langle0||1\rangle_{N,\mathrm{out}}\langle1|,
\end{aligned}
\]
and the off-diagonal part $\rho_{\mathrm{off}}^{F}$ is given by
\[
\begin{aligned}
\rho_{\mathrm{off}}^{F}&=\sum_{i,j=1(i\neq j)}^{N-1}|1\rangle_{i}\langle0||0\rangle_{j}\langle1|\bigotimes_{k=1,(k\neq i,k\neq j)}^{N-1}|0\rangle_{k}\langle0|\bigg\{[e^{-8\pi\omega(M-D)}+1]^{-1}|0\rangle_{N,\mathrm{out}}\langle0| \\
&+[e^{8\pi\omega(M-D)}+1]^{-1}|1\rangle_{N,\mathrm{out}}\langle1|\bigg\}+\sum_{i=1}^{N-1}|1\rangle_{i}\langle0|\bigotimes_{j=1(j\neq i)}^{N-1}|0\rangle_{j}\langle0|[e^{-8\pi\omega(M-D)}+1]^{-1}|0\rangle_{N,\mathrm{out}}\langle1|\\
&+\sum_{i=1}^{N-1}|0\rangle_{i}\langle1|\bigotimes_{j=1(j\neq i)}^{N-1}|0\rangle_{j}\langle0|[e^{-8\pi\omega(M-D)}+1]^{-\frac{1}{2}}|1\rangle_{N,\mathrm{out}}\langle0|.
\end{aligned}
\]
The eigenvalues of this state $\rho_{123\ldots N_{\mathrm{out}}}^{F,\mathrm{W}}$ are
\begin{eqnarray}\label{S57}
\lambda^F_{\mathrm{W}_{1}}=\frac{N-1}{N[1+e^{8\pi\omega(M-D)}]}, \quad \lambda^F_{\mathrm{W}_{2}}=\frac{1+Ne^{8\pi\omega(M-D)}}{N[1+e^{8\pi\omega(M-D)}]}.
\end{eqnarray}
The reduced density matrix $\rho_{N_{\mathrm{out}}}^{F,\mathrm{W}}$ is obtained by tracing out $N-1$ subsystems as
\begin{eqnarray}\label{S58}
\rho_{N_{\mathrm{out}}}^{F,\mathrm{W}} &=&\frac{1}{N}\bigg\{(N-1)[e^{-8\pi\omega(M-D)}+1]^{-1}|0\rangle_{N,\mathrm{out}}\langle0|\notag\\
&+&\{(N-1)[e^{8\pi\omega(M-D)}+1]^{-1}+1\}|1\rangle_{N,\mathrm{out}}\langle1|\bigg\},
\end{eqnarray}
and the eigenvalues of the density matrix above is computed as
\begin{eqnarray}\label{S59}
\lambda^{F1}_{\mathrm{W}_{1}}=\frac{(N-1)e^{8\pi\omega(M-D)}}{N[1+e^{8\pi\omega(M-D)}]}, \quad \lambda^{F1}_{\mathrm{W}_{2}}=\frac{N+e^{8\pi\omega(M-D)}}{N[1+e^{8\pi\omega(M-D)}]}.
\end{eqnarray}
Substituting these eigenvalues into Eq.(\ref{S6}), the $q$-conditional entropy
is
\begin{eqnarray}\label{S60}
S_{F,q}^{\mathrm{W},N}=\frac{1}{q-1}\bigg\{1-\frac{(N-1)^{q}+[1+Ne^{8\pi\omega(M-D)}]^{q}}{[(N-1)e^{8\pi\omega(M-D)}]^{q}+[N+e^{8\pi\omega(M-D)}]^{q}}\bigg\}.
\end{eqnarray}

\begin{figure}[htbp]
\begin{minipage}[t]{0.5\linewidth}
\centering
\includegraphics[width=3.0in,height=5.2cm]{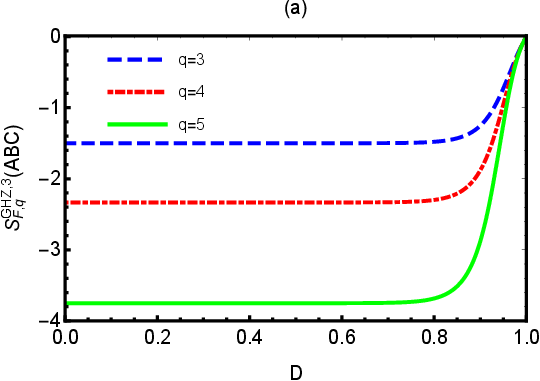}
\end{minipage}%
\begin{minipage}[t]{0.5\linewidth}
\centering
\includegraphics[width=3.0in,height=5.2cm]{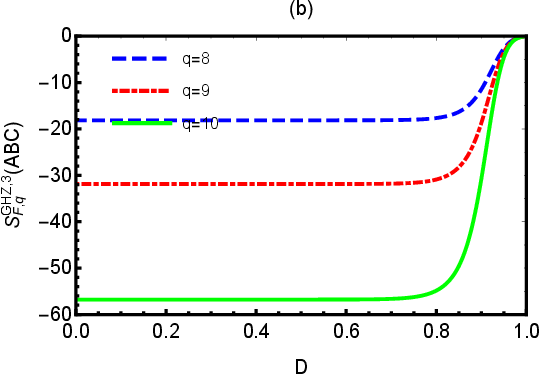}
\end{minipage}%

\begin{minipage}[t]{0.5\linewidth}
\centering
\includegraphics[width=3.0in,height=5.2cm]{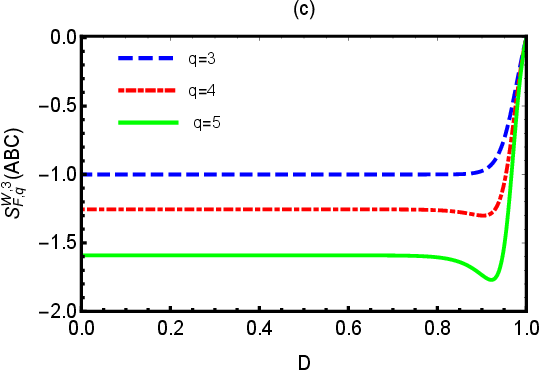}
\end{minipage}%
\begin{minipage}[t]{0.5\linewidth}
\centering
\includegraphics[width=3.0in,height=5.2cm]{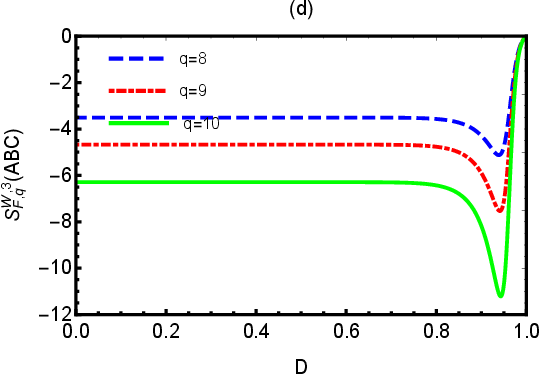}
\end{minipage}%
\caption{AR $q$-conditional entropy of GHZ and W states for fermionic field as a function of the dilaton  $D$ for various values of $q$ with $M=\omega=1$ and $N=3$.}
\label{Fig.2}
\end{figure}

In Fig.\ref{Fig.2},  we present the $q$-conditional entropy $S^{\mathrm{GHZ},_{3}}_{F,_{q}}(ABC)$ and $S^{\mathrm{W},_{3}}_{F,_{q}}(ABC)$ of fermionic field as a function of the dilaton  $D$ for different $q$, calculated from Eqs.(\ref{S54}) and (\ref{S60}).  From Fig.\ref{Fig.2}, we can see that the behavior  of $S^{\mathrm{GHZ},_{3}}_{F,_{q}}(ABC)$ in fermionic field is similar to that of $S^{\mathrm{GHZ},_{3}}_{B,_{q}}(ABC)$ of bosonic field in curved spacetime. However, a striking distinction emerges with the $q$-conditional entropy $S^{\mathrm{W},_{3}}_{F,_{q}}(ABC)$ of fermionic field, which is a non-monotonic function of the $D$.
This non-monotonicity indicates that the Hawking effect from the black hole significantly enhances the nonseparability of  W state in fermionic field. As
the $D$ increases, the initial nonseparability of  W state reaches a maximum, demonstrating that the Hawking effect induces a net increase in nonseparability.
Therefore, the properties of $S^{\mathrm{W},_{3}}_{F,_{q}}(ABC)$ of fermionic field are different from those of  bosonic field in the dilaton black hole.  Unlike the $q$-conditional entropy of bosonic field, as the $D$ approaches the mass $M$ of the black hole, the $q$-conditional entropy of fermionic field tends toward non-zero negative values for all values of $q$, indicating that both GHZ and W states retain their nonseparability in the extreme black hole limit. In other words,  at the threshold  $D\rightarrow M$, the quantum states of fermionic field still remain entangled, highlighting a fundamental difference between fermionic and bosonic fields in the context of the dilaton black hole. Furthermore, from both Fig.\ref{Fig.1} and Fig.\ref{Fig.2}, we observe that the
$q$-conditional entropy of bosonic field is consistently greater than that of fermionic field within dilaton spacetime, suggesting that the nonseparability of bosonic field is weaker than that of fermionic field under the influence of the Hawking effect. In contrast, quantum coherence is stronger for bosonic field than for fermionic field in the dilaton black hole \cite{A11,A16}, emphasizing the distinct quantum properties of these fields in a relativistic background.
Due to the differences between the Bose-Einstein and Fermi-Dirac distributions, bosons and fermions exhibit fundamentally distinct behaviors in the background of a dilaton black hole. For instance, in the extreme black hole limit, entanglement for bosons vanishes completely, whereas fermionic entanglement can partially survive, demonstrating greater resilience. As a result, fermionic nonseparability is generally more robust than that of bosons in such spacetime. The quantum coherence of multiple particles reflects the ability of each particle to maintain superposition within its local state space, while quantum entanglement refers to the non-classical correlations between particles, such that the total state cannot be factorized into a product of individual subsystems. Owing to their fundamentally different physical characteristics, it is found that fermionic coherence is consistently smaller than bosonic coherence in the dilaton black hole.
This highlights the importance of selecting the appropriate quantum resources when conducting relativistic quantum information tasks, as the choice of particle type can significantly influence the efficiency and outcome of quantum protocols.

\begin{figure}
\begin{minipage}[t]{0.5\linewidth}
\centering
\includegraphics[width=3.0in,height=5.2cm]{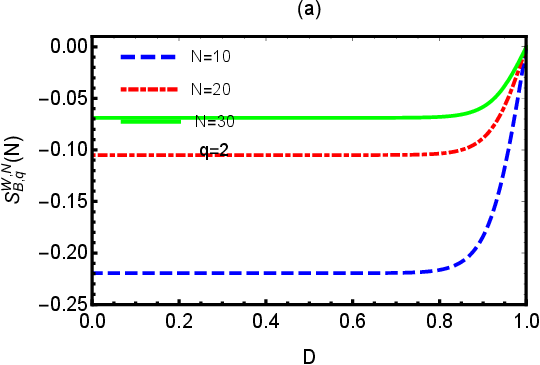}
\label{fig3a}
\end{minipage}%
\begin{minipage}[t]{0.5\linewidth}
\centering
\includegraphics[width=3.0in,height=5.2cm]{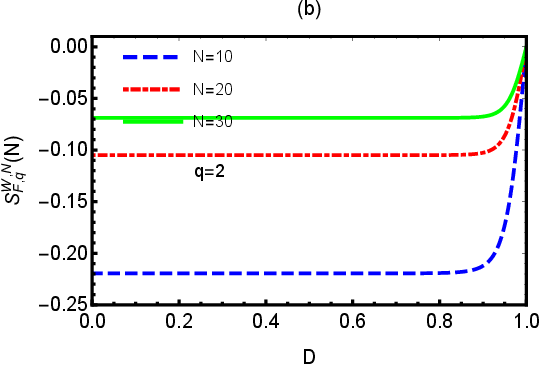}
\label{fig3b}
\end{minipage}%

\begin{minipage}[t]{0.5\linewidth}
\centering
\includegraphics[width=3.0in,height=5.2cm]{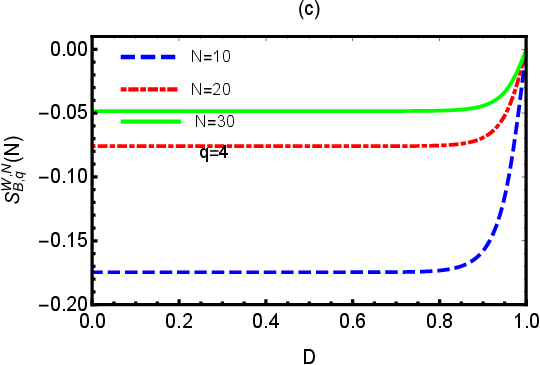}
\label{fig3c}
\end{minipage}%
\begin{minipage}[t]{0.5\linewidth}
\centering
\includegraphics[width=3.0in,height=5.2cm]{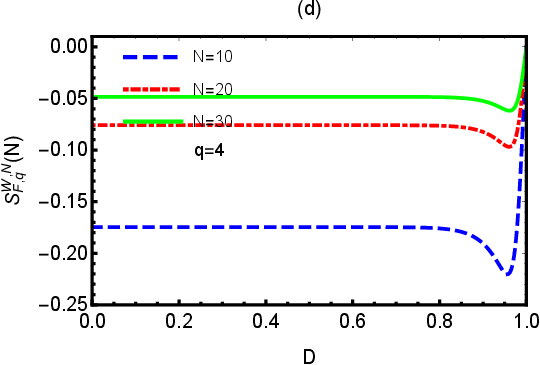}
\label{fig3d}
\end{minipage}%

\begin{minipage}[t]{0.5\linewidth}
\centering
\includegraphics[width=3.0in,height=5.2cm]{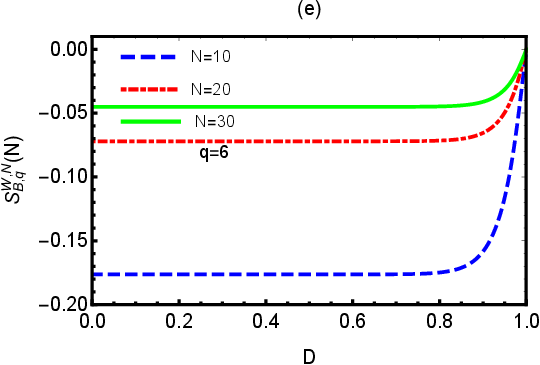}
\label{fig4a}
\end{minipage}%
\begin{minipage}[t]{0.5\linewidth}
\centering
\includegraphics[width=3.0in,height=5.2cm]{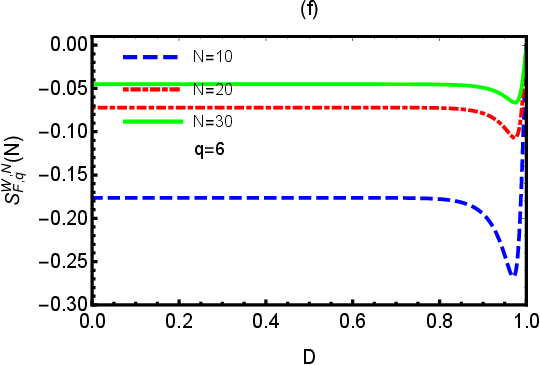}
\label{fig4b}
\end{minipage}%

\begin{minipage}[t]{0.5\linewidth}
\centering
\includegraphics[width=3.0in,height=5.2cm]{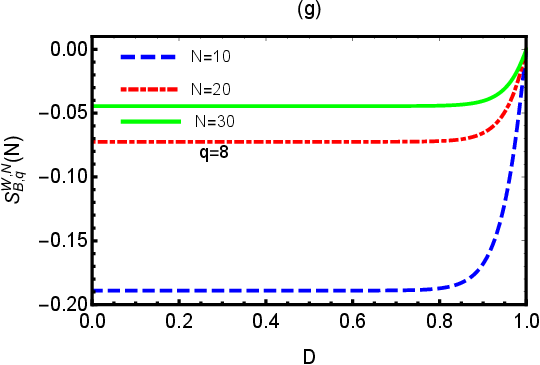}
\label{fig4c}
\end{minipage}%
\begin{minipage}[t]{0.5\linewidth}
\centering
\includegraphics[width=3.0in,height=5.2cm]{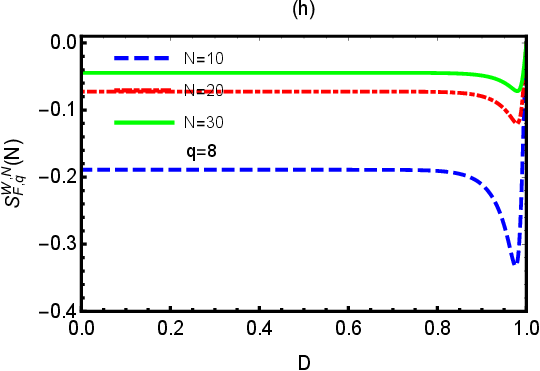}
\label{fig4d}
\end{minipage}%
\caption{AR $q$-conditional entropy of W state of both bosonic and fermionic fields as a function of the dilaton $D$ for various values of $q$ and $N$ with $M=\omega=1$.}
\label{Fig.3}
\end{figure}

Through the analytical expressions in Eqs.(\ref{S40}) and (\ref{S54}),  we find that the $q$-conditional entropy of  GHZ state remains independent of the number of qubits  $N$ in both bosonic and fermionic fields. This stands in stark contrast to the behavior observed in W state, where, as seen in  Eqs.(\ref{S48}) and (\ref{S60}), the $q$-conditional entropy is explicitly dependent on $N$ in both types of fields.
To illustrate this relationship, we plot the AR $q$-conditional entropy $S^{\mathrm{W},_{N}}_{B,_{q}}(N)$ and $S^{\mathrm{W},_{N}}_{F,_{q}}(N)$ as a function of the dilaton $D$ for different $q$ and $N$, as shown in Fig.\ref{Fig.3}. From Fig.\ref{Fig.3}, it is evident that the $q$-conditional entropy of W state increases with the growth of the $N$, indicating a decrease in nonseparability as the number of qubits grows within the dilaton black hole.  This trend is indicative of the complex interplay between quantum correlations that emerge as more qubits are introduced into the system, directly influencing the conditional entropy. This behavior is akin to the properties of quantum entanglement in W state under relativistic environments, where quantum entanglement diminishes as the $N$ grows.
Interestingly, while the nonseparability of W state weakens with increasing $N$, its quantum coherence exhibits a monotonically increasing trend within the dilaton spacetime  \cite{A16}.
This is because, as the number of particles $N$ in the W state increases, a fixed amount of quantum correlations is distributed over more subsystems, leading to a weakening of the overall entanglement (nonseparability), especially in pairwise correlations. However, since the W state involves superpositions of more basis states, its quantum coherence increases. In the background of the GHS dilaton black hole, the thermal noise induced by the Hawking effect more severely degrades entanglement, while coherence exhibits greater robustness against such noise-it can not only survive but may even increase with growing $N$.
This duality between decreasing entanglement and increasing coherence underscores the nuanced and complex nature of quantum resources in relativistic regimes. These findings emphasize the importance of carefully selecting multipartite quantum resources for relativistic quantum information tasks, as they reveal competing factors that affect the overall quantum information dynamics in curved spacetime.

\section{Conclusions}
The influence of the Hawking effect on N-partite nonseparability of GHZ and W states using the AR $q$-conditional entropy for bosonic and fermionic fields is investigated in the background of the GHS dilaton black hole.  Initially, $N$ observers are posited to share GHZ and W states in the asymptotically flat region. The scenario then evolves with $N-1$ observers remaining stationary in the asymptotic region, while a single observer hovers near the event horizon of the black hole.
We derive general analytical expressions for the N-partite $q$-conditional entropy of both bosonic and fermionic fields, which quantify the nonseparability of GHZ and W states in the dilaton black hole background.  Our findings reveal a striking difference between the behavior of bosonic and fermionic fields under the influence of  gravitational effects. Specifically, fermionic nonseparability proves to be more resilient than its bosonic counterpart, despite a weaker fermionic coherence in the dilaton black hole \cite{A11,A16}. For the bosonic field, both GHZ and W states undergo a transition from nonseparability to separability in the extreme black hole limit.
Interestingly, we observe that  the nonseparability of  GHZ state consistently exceeds that of the W state, while its quantum coherence is weaker in the dilaton black hole.
Furthermore, the nonseparability of  GHZ state remains unaffected by the number of initial particles $N$, while for  W state, the nonseparability exhibits a strong dependence on $N$. As the $N$ increases, the nonseparability of W state decreases, while its quantum coherence grows in the dilaton black hole. A particularly noteworthy result is that the Hawking effect can lead to a net increase in the nonseparability of  W state for fermionic fields, a phenomenon that contrasts with the behavior seen in bosonic fields. These results suggest that the choice of quantum resources must be tailored to different particle types and quantum state configurations when handling relativistic quantum information tasks. We hope this research not only enhances our understanding of quantum information behavior in strong gravitational fields but also aids in the effective utilization of quantum resources in future relativistic quantum information applications.

\begin{acknowledgments}
This work is supported by the National Natural
Science Foundation of China (Grant nos. 12205133 and 12175095), the
Special Fund for Basic Scientific Research of Provincial Universities in
Liaoning under Grant no. LS2024Q002, and LiaoNing Revitalization
Talents Program (XLYC2007047)
\end{acknowledgments}


\end{document}